\documentclass[]{interact}

\usepackage{epstopdf}
\usepackage[caption=false]{subfig}

\usepackage[numbers,sort&compress]{natbib}
\bibpunct[, ]{[}{]}{,}{n}{,}{,}
\renewcommand\bibfont{\fontsize{10}{12}\selectfont}

\theoremstyle{plain}

\theoremstyle{definition}

\theoremstyle{remark}

\begin{document}


\title{The two-level systems in cryogenic solids, or how to avoid stressful memories}

\author{
\name{Vassiliy Lubchenko\thanks{CONTACT Vassiliy Lubchenko. Email: vas@uh.edu} }
\affil{Departments
  of Chemistry and Physics, University of Houston, Houston, TX 77204-5003 
  \\ Texas Center for Superconductivity,
  University of Houston, Houston, TX 77204-5002}
}

\maketitle

\begin{abstract}

Structural glasses prepared by bulk quenching a liquid melt universally exhibit puzzling low-energy excitations commonly known as the ``two-level systems'' (TLSs). Recent studies indicate that ultrastable glassy films made by vapor deposition exhibit substantially fewer TLSs and, at the same time, are more stable enthalpically than conventional glasses made by quenching a melt. A similar phenomenon is observed in very stable glasses of model liquid mixtures prepared using swap Monte Carlo sampling. However, in a separate set of enthalpically stable solids, exemplified by amber matured over geological times, the two-level systems persist. In addressing this seeming conflict, we emphasize that a depletion of the TLSs, if any, means the configurational entropy of the material is lower than that of conventional glasses made by bulk-quenching a melt. 
Ageing does induce reduction in configurational entropy, but amber, we speculate, achieves enthalpic stabilization through increased bonding, not ageing. We separately comment on the discrepancy among existing predictions for the extent of cooperativity of the two-level systems. Several experiments are suggested to test the present  picture. 
\end{abstract}


\section{Introduction}

The idea of a multiverse, frequently exploited in popular culture, may seem contrived, but it is routinely realized in nature. Every solid can be thought of as a collection of universes---each universe endowed with a vibrational vacuum~\cite{BLelast}---that carries a unique historical record of its birth. This record, or one might say memory, is in the form of various geometric patterns, some of which are quite striking, such as grain boundaries or dislocations, and some that are not readily discernible. Some patterns are essentially static yet some are mobile and can be concealed by ordinary vibrations. Patterns of the striking variety are customarily referred to as ``defects'' by materials scientists, but the designation implies there is a universal, defect-free reference arrangement, something that is often hard to cleanly define in materials science and human relationships alike. One may alternatively adopt a free energy-based view of information~\cite{DayanHintonNealEtAl95}, in which patterns are therefore associated with tension---which undoubtedly helps make for a good story in any genre---yet quantifying tensions and stresses again requires one to specify an appropriate reference configuration, much like a standard state in Chemistry. Defining a standard state is, however, ambiguous because practical descriptions typically break ergodicity \cite{HLKnowledge} and, as such, will be shaped by historical precedent. 

\begin{figure}
\centering
\includegraphics[width=10cm]{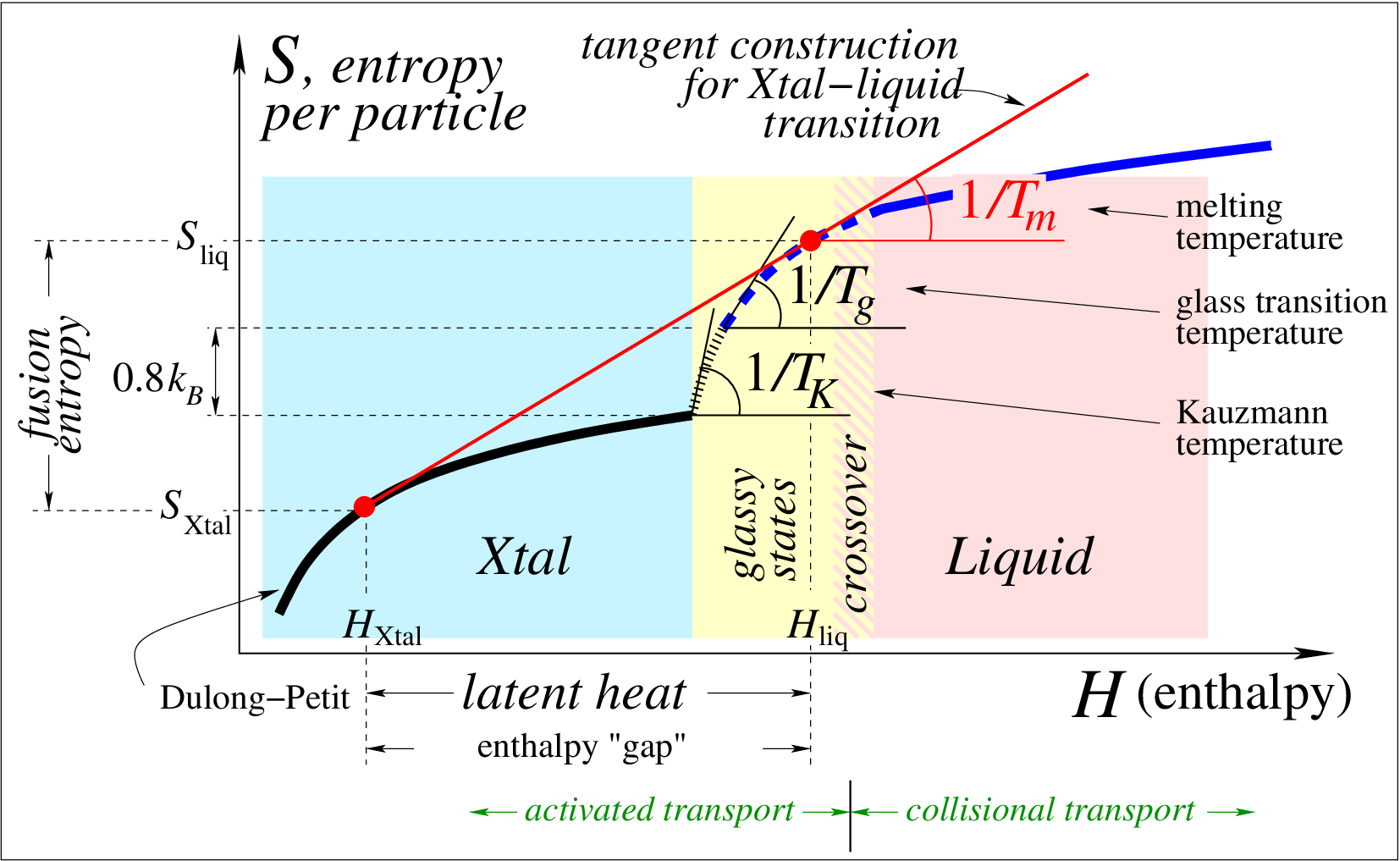}
\caption{The equilibrium equation of state (EOS) of a substance in the enthalpy range pertinent to liquid-crystal coexistence. The thick blue line depicts the liquid entropy as a function of enthalpy, the thick black line the entropy of the crystal. The solid portion of the liquid EOS corresponds to the ergodic regime $T>T_\text{cr}$, the dashed portion signifies ergodicity is broken on times less than the structural relaxation time $\tau_\alpha$. The states between $H_\text{Xtal}$ and $H_\text{liq}$ are bypassed during quasi-equilibrium crystallisation but are visited, if the liquid can be supercooled below the melting temperature. The glass transition ordinarily occurs at enthalpy values within the enthalpy gap $[H_\text{Xtal}, H_\text{liq}]$. The crossover to the landscape regime (``glassy states''), in which transport is activated, could be either above or below the melting point $T_m$, the two cases corresponding to strong and fragile liquids, respectively.} \label{SH}
\end{figure}

Fortunately, and despite being condensed assemblies of strongly interacting objects, many liquids appear to fully equilibrate at sufficiently high temperatures thus affording us a reliable reference state, independent of the sample's convoluted history; see Ref.~\cite{deng2025longtimederivationboltzmann} for a recently obtained set of rigorous results in the dilute limit. We depict this reference state by the solid blue line in Fig.~\ref{SH}. In such an equilibrium, uniform liquid, any particle will exchange places with another chemically equivalent particle so fast as to completely prevent one from labeling particles using their locations. Conversely, such labeling can be accomplished if particles are prevented from exchanging places, but the resulting breaking of ergodicity comes at a cost: It lowers the entropy by roughly $1 k_B$ per particle~\cite{MottGurneyLiquids, L_AP}. The associated free energy excess, $k_B T$ or so per particle, may seem modest yet it sometimes gives rise to remarkable effects that are strongly dependent on the detailed history of ergodicity-breaking events during freezing. 

This article addresses recent developments regarding one of these remarkable history-dependent effects. Glasses that are made by thermal quenching universally exhibit puzzling structural degrees of freedom in excess of ordinary vibrations. These strongly anharmonic, resonant degrees of freedom behave like two-level systems at sufficiently low temperatures according to bulk measurements~\cite{ZellerPohl, GG, HunklingerRaychaudhuri, AHV, Phillips, Jackle}, but experiments at the individual level indicate resonances are sometimes better described as multiple-level~\cite{Orrit}; comprehensive reviews can be found in Refs.~\cite{LW_RMP, Lrelics}. Early, semi-phenomenological treatments~\cite{AHV, Phillips} assumed the excitations come about because glasses can host strongly anharmonic, non-phononic modes owing to the lack of strict periodicity. These strongly anharmonic modes were imagined as bistable degrees of freedom each involving very few atoms---lest tunneling would be too slow---while the modes' parameters would be system-specific. Surprisingly, the TLS density of states turned out to be rigidly correlated with how strongly the TLS couple to the acoustic modes~\cite{FreemanAnderson}, suggesting these mysterious degrees of freedom might be, in fact, inherent to disordered solids~\cite{YuLeggett, Lrelics} and quenched glasses, in particular~\cite{LW}. The latter scenarios explicitly require that the TLSs be truly collective modes that involve many atoms~\cite{Lrelics}. Subsequent experimental work on a variety of {\it non}-glassy disordered solids showed~\cite{Pohl_review} that degrees of freedom in excess of Debye-type vibrations are indeed common when long range order is absent, but the amount of these modes is much lower than in quenched glasses proper, nor are their properties universal. 

We associate glasses with highly metastable configurations of atoms that are characterized by an excess density of structural states, the latter manifesting as the TLSs themselves or, to give another famous example, as the Boson Peak~\cite{LW_BP, LW_RMP}. Conversely, the point symmetry of many periodic structures is often quite compatible with local bonding rules, thus implying relatively low enthalpy while, at the same time, making non-phononic degrees of freedom very costly. This suggests a general link between the enthalpic stability of a solid, on the one hand, and the presence of non-phononic degrees of freedom, on the other hand. A recent body of experimental and computational work, which we review in Section~\ref{review}, indicates however that enthalpically stabilizing a glassy solid may or may not lead to a decrease in the TLS density of states. Indeed, ultrastable glassy films~\cite{Moratalla2023} and deeply supercooled model liquids~\cite{MocanuNTLS} alike show a substantial decrease in the number of TLSs, whereas amber matured over geological times does not~\cite{PhysRevLett.112.165901}. In Section~\ref{discussion}, we attempt to resolve this seeming conflict by showing that the puzzling diversity of observations in those stable solids stems from distinct ways those solids break ergodicity. Finally in Section~\ref{summary}, we summarize and suggest a number of new experiments and ideas for simulation that could further test this picture.


\section{When more stability implies fewer two-level systems, and when it does not: A review of recent experimental and computational studies}

\label{review}

An instructive preparation protocol for making a glass is to first quench a liquid melt quickly below the optimal crystallization window, the latter determined by the interplay of the crystal nucleation barrier and the viscosity, respectively. The liquid is now quite viscous but it still equilibrates on laboratory timescales; we will assume for concreteness there is no detectable crystallinity. Further quenching will cause the melt to become even more viscous and to eventually fall out of equilibrium, when the relaxation rates of the slowest processes become lower than the rate of the quench. Alongside, the heat capacity of the liquid will suffer a pronounced drop within a relatively narrow temperature interval (up to several ten Kelvins in width)~\cite{AngellScience1995}, implying a subset of degrees of freedom have frozen out. A ``glass transition'' is said to have occurred. In contrast with crystallization, this ``transition'' is continuous in that it presents no latent heat or abrupt changes in other properties, yet there is no identifiable criticality either. Slower quenches result in lower glass transition temperatures and produce glasses that are more stable, i.e., exhibit lower enthalpy. Faster (thermal) quenches, on the other hand, not only imply less stability overall, but often result in built-in stresses due to variations in the rate of cooling and contraction across the sample. Dramatically exemplified by Prince Rupert's drops, this type of built-in stress, which varies on {\em macro}scopic lengthscales, is nonetheless low at the molecular level and will not be considered here.  

Bulk-quenched glasses themselves appear to offer relatively little flexibility in terms of preparation protocol because the properties of the so made solid depend on the quenching rate logarithmically weakly~\cite{LRactivated}. The range of rates practically achievable by thermal quenching is determined by the thermal conductivity of the substance and is not particularly broad, in the first place. One may expect that the density of states for the TLSs should vary by about an order of magnitude within the range of quench rates available in principle, the more leisurely made glasses predicted to exhibit fewer resonances~\cite{LRactivated, LW_RMP, Lrelics}. 

\begin{figure}
\centering
\subfloat[Low-temperature heat capacity of four distinct types of solid made of the same compound~\cite{Moratalla2023}, the types labeled in the legend. ``USG'' refers to ''ultrastable glass,'' ``iso'' to ``isotropic,'' and ``ani'' to  ``anisotropic''. The temperatures at which the ultastable films were deposited are also indicated. The $0.85 T_g$ film is the more stable of the two. Reprinted from Ref.~\cite{Moratalla2023} under a Creative Commons CC BY 4.0 International License.]{%
\resizebox*{7cm}{!}  {\includegraphics{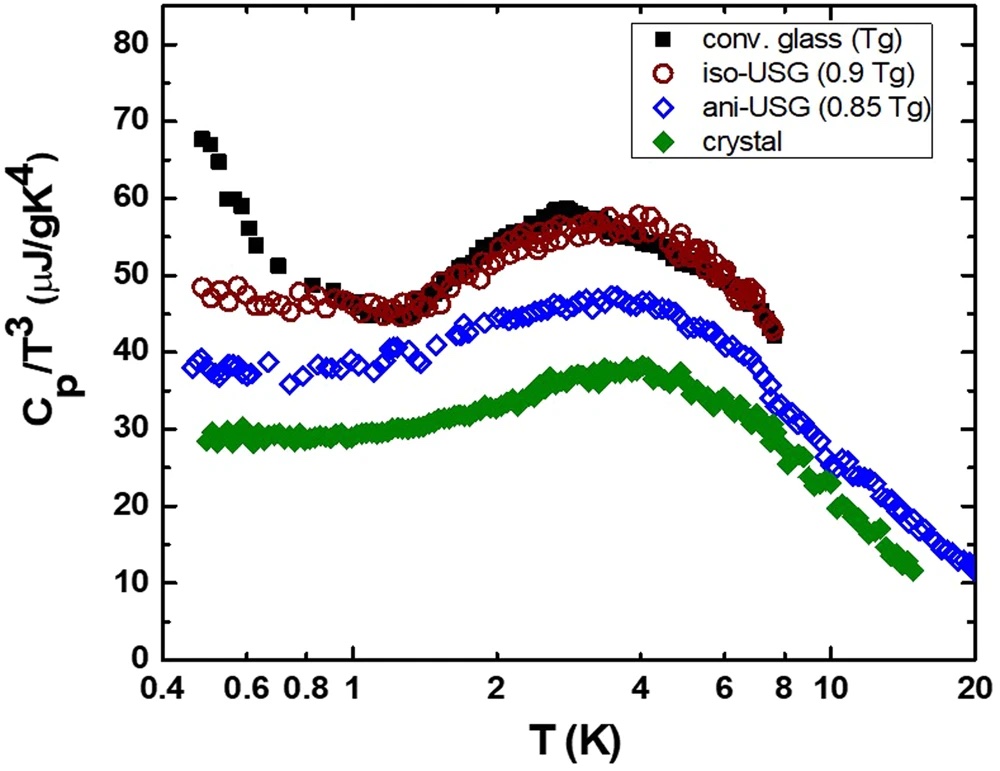}}}\hspace{5pt}
\subfloat[Density of the detected two-level systems in ternary Lennard-Jones (TLJ) and polydisperse soft-sphere (PSS) glasses as a function of stability of the sample, as reflected in the value of the activation barrier for $\alpha$-relaxation~\cite{MocanuNTLS}. Reprinted from Ref.~\cite{MocanuNTLS} by permission.]{%
\resizebox*{7cm}{!}{\includegraphics{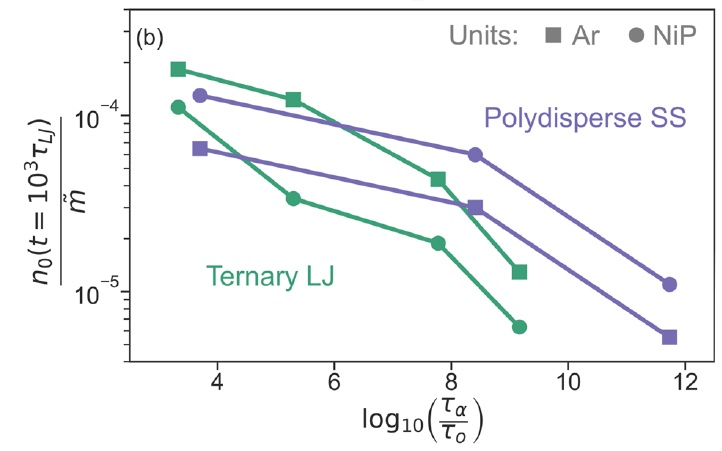}}}
\caption{Signatures of depletion in the number of two-level systems in (a) ultrastable glasses~\cite{Moratalla2023} and (b) model Lennard-Jones mixtures~\cite{MocanuNTLS}.} \label{TLSexp}
\end{figure}

It turns out glassy solids come about in other ways, too. In an exciting development, Ediger and coworkers discovered, some time ago, that vapour deposition at temperatures around 85\% of the laboratory glass transition temperature $T_g$ produce glass-like films that are exceptionally stable ~\cite{EdigerScience2007}; these materials are often referred to as ``ultrastable glasses.'' More recently, Ramos and coworkers~\cite{PerezCastaneda05082014, Moratalla2023} established that in ultrastable glasses, there is little in the way of the (near) linear heat capacity we associate with two-level systems, at accessed temperatures. In Fig.~\ref{TLSexp}(a), we show the most recent measurements, due to Moratalla et al.~\cite{Moratalla2023}. Fitting ambiguities aside, the data on the most stable film could be consistent with a complete lack of TLSs, while the film of the intermediate stability shows only a small uptick in the density of states at low-energy side of the accessed range, significantly smaller than in the bulk-quenched glass.

Meanwhile, not all is quiet on the computational front: Increasingly more efficient algorithms have been developed for Gibbs sampling of liquid mixtures, that allow for two differently sized particles to swap places, see Ref.~\cite{PhysRevX.7.021039} and references therein. This is enabling one to obtain equilibrium configurations for model liquids at such low temperatures where sampling via physical-like moves is impractically slow~\cite{doi:10.1073/pnas.1706860114}.(In physical terms the latter slowness corresponds with very high viscosities.) More recently, a tour-de-force computational study due to Mocanu et al.~\cite{MocanuNTLS} has addressed the question of the quantity of TLS-like modes in highly stable, model glassy solids prepared using a combination of a swap Monte Carlo algorithm and conventional molecular dynamics. Hereby Mocanu et al.~\cite{MocanuNTLS} explored potential energy surfaces for several stable glass-like samples of this type, with the aim of identifying bistable energy profiles. The corresponding (multi-particle) motions are candidate modes that could give rise to low-energy resonances. The latter authors have established that search for candidate TLSs has a substantially lower success rate for more stable structures, see Fig.~\ref{TLSexp}(b). The amount by which the success rate is lowered is quite consistent with predictions~\cite{LW, LW_RMP, LRactivated} based on the RFOT theory~\cite{LW_ARPC, L_AP}, but the observed amount of cooperativity of the candidate motions is substantially lower than what Ref.~\cite{LW} predicts. The energy trend of the cooperativity is in seeming conflict with the RFOT theory, too.

Still, higher stability does not necessarily imply fewer TLSs. P\'erez-Casta\~neda et al.~\cite{PhysRevLett.112.165901} have determined that amber matured over geologically long times is substantially more stable than its rejuvenated counterpart, the degree of extra stabilization comparable to that found in ultrastable glass films. Yet the TLS contribution to the low-$T$ heat capacity was unaffected; it also stayed put after annealing and re-vitrifying the samples. The latter notion is significant because its glassy properties notwithstanding, amber is not an ordinary glass. Various chemical processes other than translations and vibrations take place during its formation, such as polymerization and cross-linking, nor can one know how the conditions varied as the sample was maturing over the aeons.  

\section{Discussion}

\label{discussion}

The three sets of findings described above can be profitably discussed within the framework provided by the random first order transition (RFOT) theory, see Refs.~\cite{LW_ARPC, L_AP} for reviews, but some of the forthcoming conclusions are based on entirely generally grounds. 

\subsection{Breaking ergodicity while making a memory: Free energy cost and the residual stress}

\label{pedagogical}

\begin{figure}
\centering
\includegraphics[width=7cm]{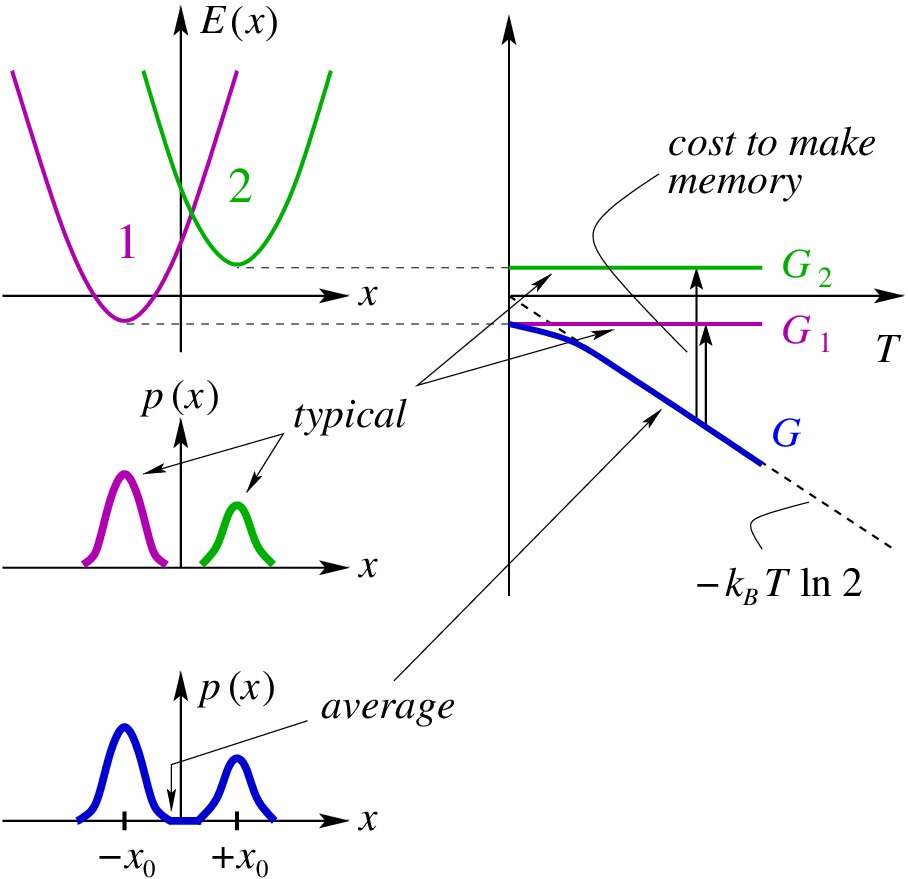}
\caption{Graphical explanation of ergodicity breaking. $p(x)$ stands for the probability distribution of the variable $x$. Note that we have not included the additive contribution $-T s_\text{vib}$ of the vibrational entropy to the free energies $G_1$, $G_2$, and $G$, for clarity. This contribution is the same for all three free energies.} \label{memory}
\end{figure}

We begin with just such a general notion. Ergodicity breaking is a key component of the physics of glasses that has many observable consequences. Some of its statistical aspects are, however, rather subtle, so let us illustrate them using an elementary example; this example is of some merit on its own. Consider a potential energy that operates on two degrees of freedom, one continuous, $-\infty < x < +\infty$, and one discrete, Ising spin-like $\sigma = \pm 1$:
\begin{equation} \label{Exs}
    E(x, \sigma) = \frac{k}{2}(x -  \sigma \, x_0)^2  + \varepsilon \sigma + E_0,
\end{equation}
where $x_0 > 0$ and $0 < \varepsilon < k x_0^2$ by construction. The constant $E_0$ specifies the energy reference and, thus, depends on the preparation history. As a function of $x$, $E(x, \sigma)$ simply amounts to two identical replicas of a parabola. The bottom of each parabola corresponds to a vibrational minimum. Despite its simplicity, the energy function~(\ref{Exs}) is a useful representation of the two states of a single-bit memory unit, such as a ferromagnetic domain. The forthcoming statements are graphically summarized in Fig.~\ref{memory}. 

Suppose that the activation energy for inter-minimum transitions is significantly greater than temperature, so that the characteristic rate $1/\tau_c$ for these transitions is much lower than the rate of vibrational relaxation $1/\tau_\text{vib}$. On times such that $\tau_\text{vib} < t < \tau_c$, the vibrational degree $x$ of freedom is equilibrated, but the configurational degree of freedom $\sigma$ is not. Thus, ergodicity is transiently broken on times $t < \tau_c$. Alternatively said, the memory unit can reliably store information, i.e. the record of its preparation history, on times $t < \tau_c$. In equilibrium, the average value of $x$ is close to midway between the two minima, for $\varepsilon/k_B T$ small, and not at all representative of the actual behaviour of the variable $x$. In fact, the system is strictly unstable at $x = \langle x \rangle$, because both parabolas are sloped there. Instead, typical values of $x$ are clustered around individual minima, of which there are two, resulting in a typical free energy that is two-valued. On the other hand, the average---that is, equilibrium---free energy is unique, of course, and always lower than the lesser of the two typical free energies. Indeed, the free energy of the $\sigma = -1$ (``memory 1'') configuration is $G_1 = - \varepsilon + E_0 - T s_\text{vib}$, where $s_\text{vib}$ is the vibrational entropy corresponding to an individual parabola. Likewise for $\sigma = +1$ (``memory 2''), $G_2 = + \varepsilon + E_0 - T s_\text{vib}$. The equilibrium free energy is $G = E_0 - k_B T \ln \left[ \left(e^{\varepsilon/k_B T} + e^{-\varepsilon/k_B T} \right)/2 \right] - T (s_\text{vib} + k_B \ln 2)$. We see that there is a lowering of the equilibrium free energy, relative to the free energies of individual minima, that tends to $- T (k_B \ln 2)$ for small $\varepsilon/k_B T$. The $2$ under the logarithm stems from the number of the minima being $2$, of course. The temperature scaling of $- T (k_B \ln 2)$ indicates that this free-energy lowering is of entropic nature. Thus one may designate $s_c = k_B \ln 2$ as the configurational entropy, the word ``configuration'' referring to an individual vibrational minimum. The full entropy is, then, $s = s_\text{vib} + s_c$.

The free energy cost of making a memory 1 is given by $(G_1 - G) > 0$ and for memory 2 it is $(G_2 - G) > 0$. These are proper free energy costs that are independent of the arbitrary constant $E_0$ and quantify probabilities in a history-independent fashion. Incidentally, note that when so properly defined, the norms for the two ``typical'' distributions should add together to give the norm of the equilibrium distribution.  As already alluded to, the free energy cost $(G_i - G) \rightarrow T s_c$ for small $\varepsilon$. According to Fig.~\ref{memory}, the cost of making the more stable memory vanishes only at $T=0$. Equivalently, the quantity $-T s_c$ can be thought of as the free energy stabilization due to restoring  ergodicity. The thermal availability of both configurations 1 and 2 comes down to the question whether the free energy cost of making memory 2 can be compensated, within a thermal energy scale $k_B T$, by the free energy gain of restoring ergodicity if initially stuck in memory 1, and vice versa for the $2 \rightarrow 1$ process. 

The equilibrium free energy $G$ provides a proper reference energy to define probabilities of individual vibrational minima in a way that is independent of the preparation protocol. (This is entirely analogous to why we need standard states in Chemistry.) $G$ is unambiguously determined through calorimetry, by cooling from an ergodic state, yet has the peculiar property that there are essentially no actual configurations of the variable $x$ associated with it. Instead, the equilibrium free energy $G$ corresponds to a statistical mixture of configurations that are not clustered within a single compact region in the phase space, but, instead, belong to {\em two} compact regions that are essentially disconnected: The sampling rate of intermediate regions decreases exponentially fast as the temperature is lowered. This state of affairs is a direct consequence of the ergodicity breaking, even though the ergodicity breaking is transient. Transient or not, ergodicity breaking implies that the ``average'' can be no longer conflated with a ``typical,'' the latter not being well-defined in the first place.  

The elementary notions above are helpful in thinking about the transient ergodicity breaking in glassy melts, which takes place when the fluid is cooled below the dynamical crossover at temperature $T = T_\text{cr}$,  whereby transport becomes activated. The activated-transport regime is indicated using the broken portion of the thick blue curve in Fig.~\ref{SH}, the breaks meant to emphasize that the melt breaks ergodicity on times less than the $\alpha$-relaxation time $\tau_\alpha$. The alpha relaxation time plays the role of the configurational relaxation time $\tau_c$ of the memory unit discussed earlier. Whereas above $T_\text{cr}$ the liquid is spatially uniform and ergodic on all relevant time scales, the long-time averaged profile of the liquid density {\em below} $T_\text{cr}$ is still spatially uniform, but {\em typical} density profiles are now each an aperiodic collection of sharp peaks each centered at the vibrationally averaged location of an individual particle~\cite{dens_F1}. There are $e^{s_c N}$ distinct aperiodic collections like that for a regions of size $N$, where $s_c$ is the configurational entropy per particle.  Consequently, the average value of the free energy of a region of size $N$ is lower than a typical value by the amount $T s_c N$, because the system typically resides in a particular vibrational minimum. The dashed blue line in Fig.~\ref{SH} can be regarded as a smooth analytic continuation from the fully ergodic regime at $T>T_\text{cr}$, but it is actually a sum of two entropies, the vibrational entropy $s_\text{vib}$ and the configurational entropy $s_c$, respectively. Consequently, one may determine $s_c$ by subtracting the vibrational entropy from the full entropy, the latter determined by calorimetry. The vibrational entropy of the melt or glass is difficult to determine accurately, because of a preponderance of various anharmonic modes, but we expect it to be numerically close to that of the corresponding crystal. 

The state-counting argument in Ref.~\cite{LW_aging} yields the smallest region size, $N^*$ particles, that can typically reconfigure in an equilibrated melt. At this special region size, the mismatch penalty $\Gamma(N)$ due to replacing the current (aperiodic) structure by an equivalent aperiodic structure must be compensated, within the thermal energy scale $k_B T$, by the entropic stabilization $-T s_c N$ due to restoration of ergodicity. Thus the region is guaranteed to have an alternative, yet typical aperiodic configuration. Furthermore, the mismatch penalty $\Gamma(N)$ happens to be equal to the magnitude of a typical fluctuation of the free-energy of an individual vibrational minimum for a region of size $N$~\cite{LRactivated}. The fluctuations are much slower than the vibrations and, thus, represent residual stress that is frozen-in on the vibrational timescale. 

The reference value for the residual stress in a glassy melt is, therefore, the average value of the free energy---even though there are essentially no actual configurations associated with this energy reference! This subtlety is shared by all systems that break ergodicity. For instance, the chemical potential, which we associate with the cost of adding a particle into the system, loses this literal meaning in solids. It is most straightforward to consider periodic crystals. Inserting a particle into a mechanically stable crystal or removing a particle from such a crystal leads to a non-representative, costly configuration, such as an interstitial or vacancy. Instead, particles ordinarily join crystals through a surface, which is not a bulk process. Consistent with these notions, distinct crystal faces melt at different temperatures~\cite{L_Lindemann}. Nonetheless, one may still unambiguously assign thermodynamic potentials to a crystalline solid through calorimetry, i.e, effectively by analytic continuation starting from an ergodic state. This is also a good place to reiterate that the crystal state, if any, cannot be a reference state for residual stresses because it corresponds with an entirely different portion of the phase space separated by a discontinuous transition from the melt. In other words, a glass---molten or frozen---may not be thought of as a defected crystal.

\subsection{Ultrastable films and local ordering}

\begin{figure}
\centering
\includegraphics[width=10cm]{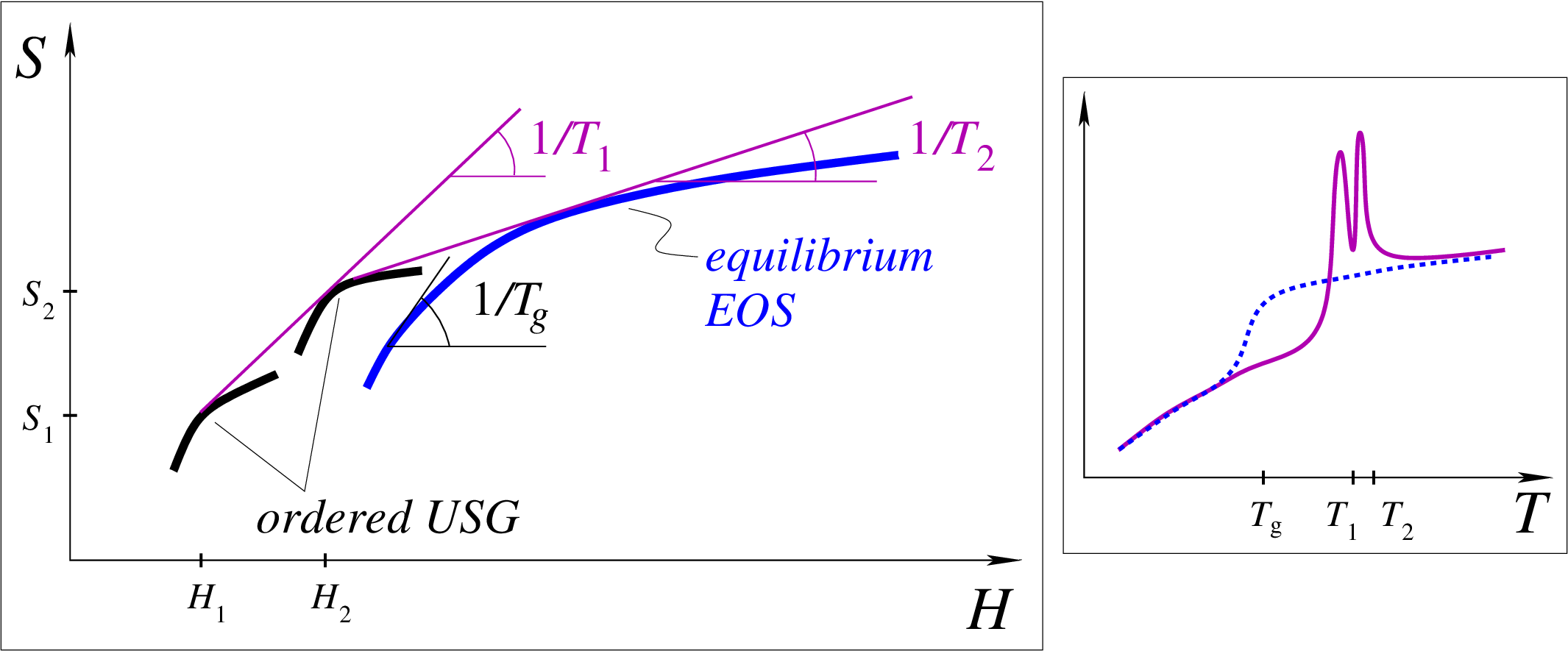}
\caption{Main graph (left): Sketch of the enthalpy dependence of the entropy for a liquid that allows for two low-enthalpy phases that exhibit local ordering; these two phases are labelled ``ordered'' USG. ``USG''=``ultrastable glass''. ``EOS''=``equilibrium equation of state.'' The corresponding down-scan (blue dashed) and up-scan for DSC are sketched in the auxiliary graph on the right. Note the respective slopes of the two double-tangent lines in the main graph should be much closer to each other, but were drawn this way to avoid cluttering.} \label{SEseries}
\end{figure}

From the very beginning~\cite{EdigerScience2007}, the differential scanning calorimetry (DSC) of ultrastable glass films indicated that the films are different from bulk-quenched glasses: The melting peak often displays a substructure in the form of a set of relatively sharp peaks, sometimes flanked by shoulders. The samples in study \cite{PerezCastaneda05082014} exhibit melting peaks that are sharp indeed; some are as sharp as the peak for the crystal-to-liquid transition. These sharp peaks indicate that the melting of the ultrastable film is accompanied by a series of phase transitions, each transition exhibiting a latent heat. (The total amount of this latent heat is smaller than the latent heat of crystal melting.) To a discrete entropy change at constant temperature there corresponds a discrete change in enthalpy. Thus it is possible that the aforementioned series of phase transitions corresponds to a density of states that consists of a set of separate blocks as depicted in Fig.~\ref{SEseries}. When drawing Fig.~\ref{SEseries}, we assumed there were two DSC peaks for concreteness. Note the samples in Ref.~\cite{Moratalla2023} exhibit one sharp peak and/or an extensive shoulder, the latter caused by cracking, pointing toward a discontinuous transition, too. These observations are consistent with the conclusions of a recent computational study~\cite{doi:10.1073/pnas.2220824120} that ultrastable films melt via nucleation. In contrast, a quenched glass that has not aged much would melt by gradual softening.

Ultrastable films commonly exhibit a varying degree of local ordering and anisotropy, even though they lack long range order. The more stable film in Ref.~\cite{Moratalla2023} exhibited few, if any, TLS, while displaying more ordering than the less stable film. The melting peaks observed during de-vitrification of ultrastable films can thus be ascribed to transitions between distinct types of local ordering and, eventually, to the equilibrium melt; these transitions are indicated by the tangent lines in Fig.~\ref{SEseries}. Ordering is favoured enthalpically, if it enables better packing. And, indeed, ultrastable films are consistently higher density than the isotropic glass. The kinetic accessibility of these densely packed aperiodic phases, during vapour deposition, can be traced to the remarkable fact that surface diffusion  can be enhanced by many orders of magnitude relative to the bulk~\cite{doi:10.1021/acs.jpcb.1c01739, stevenson:234514}. In any event, the double tangents in Fig.~\ref{SEseries} emphasize that for discontinuous transitions, the enthalpic stabilization and the entropic stabilization, respectively, go hand in hand; one cannot have one without the other.

The findings in Refs.~\cite{PerezCastaneda05082014, Moratalla2023} are qualitatively consistent with the RFOT theory. In the simplest approximation~\cite{LW, LW_RMP}, the density of states of the TLSs in a quenched glass is determined by the glass transition temperature $T_g$ and the typical volume $\xi^3$ that reconfigures during activated transport in the melt just above the glass transition:
\begin{equation} \label{nTLS}
    n_\text{TLS} \simeq \frac{1}{T_g \, \xi^3}.
\end{equation}
In absolute terms, the length $\xi$ is a couple of nanometres in a melt just above the laboratory glass transition~\cite{RWLbarrier, LRactivated}, but will decrease with heating. More instructive is the view of the reconfiguring region as a collection of relatively rigid, chemically equivalent molecular units, which we conventionally call ``the beads''~\cite{LW_soft}. This view is straightforwardly tied to actual substances by calibrating the fusion entropy of the material to the fusion entropy of a reference monatomic substance with weak cohesive interactions, such as argon. This is equivalent to introducing an isotropic ultraviolet cut-off in the theory~\cite{BL_6Spin, BLelast}. The volume of a reconfiguring region, if expressed in terms of the bead volume $a^3$ becomes a dimensionless number that characterizes the degree of cooperativity during activated transport:
\begin{equation} \label{xiNa}
    \xi^3 \equiv N^* a^3.
\end{equation}
The cooperativity size $N^*$ is a couple hundred near the laboratory glass transition---subject to system-specific variations---but decreases with heating to a couple dozen at the dynamical crossover~\cite{LW_soft}, at $T=T_\text{cr} > T_g$, above which the vibrational and translational motions can no longer be cleanly separated. The dimensionless number $N^*$ is convenient because it is tied to other important dimensionless characteristics of an equilibrated melt, i.e., the $\alpha$-relaxation barrier in thermal units and the configurational entropy per bead~\cite{L_AP}: 
\begin{equation} \label{FNsc}
    F^\ddagger/k_B T \simeq 2.6 (N^*)^{1/2} \simeq 32./s_c,
\end{equation}
thus allowing one to draw comparisons across a wide range of chemically distinct substances and melting temperatures. 

In practical terms, ergodicity breaking means equilibration is slow because the system becomes kinetically arrested in small segments of the full phase space. In an equilibrated glassy melt, these segments correspond to free energy minima. The cooperativity size $N^*$ corresponds to the smallest region that can equilibrate. The number of configurations available to this region is $e^{s_c N^*/k_B}$, where the reader will recognize the quantity $s_c N^*$ as the complexity of the reconfiguring region~\cite{Capaccioli, Berthier}. According to Eq.~(186) of Ref.~\cite{L_AP}, the complexity happens to be simply related to the activation barrier $F^\ddagger$: $s_c N^*/k_B = 4 F^\ddagger/k_B T$. The fraction of the full space phase that can be sampled in a fixed time can be estimated as the rate of escape $e^{-F^\ddagger/k_B T}$ times the fraction of the phase space $e^{-s_c N^*/k_B}$ occupied by a single configuration, to yield $\sim e^{-5 F^\ddagger/k_B T} \simeq e^{- \text{const}/s_c}$. In other words, smaller values of the configurational entropy correspond to more ergodicity breaking. At the same time, the total residual stress in the sample also decreases. Indeed, since the residual stress matches the free energy gain due to restoration of ergodicity, $\Gamma(N^*) = T s_c N^*$, the total residual stress per particle is $\Gamma(N^*)(N/N*)/N = T s_c$. Only in the putative ideal glass, \`a la Kauzmann~\cite{Kauzmann}, $s_c = 0$, would there be no residual stress caused by ergodicity breaking and no two-level systems, by Eqs.~(\ref{nTLS})-(\ref{FNsc}). 

Thus the spatial component of the density of states of the two-level systems is directly tied to the entropic content of the melt near the glass transition and, in turn, to the extent of ergodicity breaking and amount of residual stress: Less configurational entropy at freezing $\Rightarrow$ more ergodicity breaking, less residual stress, fewer TLSs. The layering within an ultrastable film implies there are correlations extending along the layer. This effectively amounts to the bead becoming anisotropic while, also, becoming bigger along two spatial dimensions. Increasing the bead volume---while keeping kinetic rates steady!---translates, per Eqs.~(\ref{nTLS})-(\ref{FNsc}), into larger cooperativity volumes and, hence, a lower density of the TLS. Since the length $\xi$ also provides the characteristic length-scale for spatial distribution of residual stress, one may likewise link depletion of the TLSs to a decrease in the amount of residual stress. 

One may contrast the above scenario with another possibility, in which the ultrastable films, instead, would belong on the equilibrium equation of state (EOS) line, depicted by the thick blue line in Figs.~\ref{SEseries} and \ref{SH}. The latter possibility would be hard to reconcile with the sharp DSC peaks seen during the melting of ultrastable films, but still it seems instructive to consider it. Although the melting temperature of the film is higher than the laboratory glass transition, the {\em equilibrium} melting temperature does not have to be, since discontinuous transitions often overshoot. This uncertainty aside, Fig.~\ref{SEseries} indicates that for the same amount of enthalpic stabilization, the entropy stabilization following a discontinuous transition from the melt to the film is smaller than if the sample followed the equilibrium equation of state; this is state 2 in Fig.~\ref{SEseries}. In the equilibrium case, the entropy decrease should be about 50\%~\cite{stevenson:234514}. This would imply a decrease in the TLS density of states by a factor of 4 or so, by Eqs.~(\ref{nTLS})-(\ref{FNsc}). Moratalla et al.~\cite{Moratalla2023} find a greater extent of depletion of the TLSs in their most stable film. Thus the configurational entropy in the ordering scenario is less than in the EOS picture, while the full entropy may be greater. If so, the vibrational entropy is greater, too. This is quite reasonable since a better packed structure should have less steric hindrance and, hence, lower vibrational frequencies. There is also the possibility (state 1 in Fig.~\ref{SEseries}), in which both the enthalpy and entropy of the film are less than in the equilibrium melt. Still, it seems likely that the vibrational entropy of the film should be greater than in the equilibrium melt.

\subsection{Particle swaps as a way to stabilize mixtures, while reducing the density of states}

Mocanu et al.~\cite{MocanuNTLS} use an extrapolation to determine the dynamical range accessible to their swap Monte Carlo algorithm and arrive at a figure close to nine orders of magnitude. This figure approaches the dynamical range, eleven-to-twelve orders of magnitude, traversed by actual substances upon cooling from the crossover temperature $T_\text{cr}$ down to the glass transition temperature $T_g$. The configurational entropy decreases, upon cooling from $T_\text{cr}$ to $T_g$ \cite{RL_Tcr}: $\Delta s_c = 0.8 k_B - 1.75 k_B \approx -1.0 k_B$ per bead. For slower quenches, this entropy drop would be larger, perhaps even as large as $\approx -1.75 k_B$, for $T_g$'s approaching the putative Kauzmann temperature $T_K$. This entropy loss is greater than the Mott and Gurney estimate~\cite{MottGurneyLiquids},  but there is no contradiction here: Entropy depends on the volume; some of the decrease of the configurational entropy with cooling is caused by thermal contraction.

The stabilization-induced decrease in the success rate of finding TLS-like modes in Ref.~\cite{MocanuNTLS}, Fig.~\ref{TLSexp}(b), is consistent with predictions of the RFOT theory. Distinct, RFOT-theory based approximations~\cite{XW, LW_soft, RWLbarrier, LRactivated} each predict that the cooperativity size $N^*$ should increase by a factor of 10 or so, going from $T_\text{cr}$ to $T_g$. This corresponds to an order of magnitude fewer TLS, per Eqs.~(\ref{nTLS})-(\ref{FNsc}). 

However, the RFOT theory also predicts that the volume $\xi^3$ not only determines the spatial part of the density of states, as in Eq.~(\ref{nTLS}), but it also gives a lower bound on the typical size of a region that reconfigures during a tunnelling event. The essential physics is this~\cite{LW}: Since a region of volume $\xi^3$ can rearrange in equilibrium, it is guaranteed to have an alternative energy minimum. But the lowest available transition state energy to reconfigure at this size would be too large for tunnelling to proceed. Larger regions, on the other hand, will have exponentially more configurations in resonance with any given value of energy. Already at volume $\simeq 1.1 \cdot \xi^3$, at least one transition state can be found that is in sufficient resonance with the initial and final state of a candidate tunnelling mode. This enables tunnelling to proceed, consistent with the fact that the nuclei need not move beyond a typical vibrational amplitude. Since these low-barrier, cooperative modes correspond to resonant tunnelling, they become damped at all temperatures above the characteristic energy scale of the Boson peak, the latter being a fraction of the Debye temperature~\cite{LW, LW_Wiley}. 

The large amount of cooperativity during tunneling predicted in Ref.~\cite{LW} may seem counter-intuitive, but established examples of similarly cooperative tunnelling events are well known. For instance, the Fr\"ohlich~\cite{RevModPhys.63.63} polaron or the Peierls-Brazovsky~\cite{ISI:A1981MD41000002, RevModPhys.60.781} polaron can travel at vanishing temperatures, {\em while the lattice vibrations remain in their ground state}. Each of the tunneling events of a TLS is analogous to a polaron hopping over hundred sites or so, the sites situated within a compact region that is only 5-6 sites across. The polaron example is not simply an analogy. We have quantitatively argued that deep midgap electronic states~\cite{ZL_JCP, ZLMicro2, LL2} and polarons~\cite{LL2, KLpolaron} in a specially family of glasses, viz. amorphous chalcogenide alloys, reside on transition-state configurations for $\alpha$-relaxations. The lack of pinning of the Peierls-Brazovsky polaron comes about because once formed, the polaron is accompanied by several contracted bonds and an exactly matching number of stretched bonds. Indeed, the polaron is bound to a pair of two topological defects of opposite polarity, corresponding to local excess and deficit of bonding, respectively~\cite{RevModPhys.60.781, KLpolaron}. Consequently, the contraction of a subset of bonds is exactly matched by dilation of the rest of the bonds---as the polaron marches on---resulting in little-to-no energy variation. This type of bond-switching mechanism was proposed by Grotthus already in 1806 to explain the high mobility of protons in water~\cite{CUKIERMAN2006876}. Just because the mechanism was proposed so long ago does not mean it is obvious or easy to visualize (except in 1D). The chalcogenides represent a felicitous situation, where the lattice stress has a very clear electronic signature that is of topological nature. Indeed, the electronic wave-function shows a remarkable amount of de-localization that would be difficult to reconcile with how deep the state is inside the forbidden gap~\cite{LL2}. Amusingly, the topological midgap electronic states seem to be a physical realization of Witten's fermions~\cite{witten1982}. Now in Lennard-Jones mixtures, there are no bonds per se, let alone electrons, yet this is precisely what the argument in Ref.~\cite{LW} accomplishes: It shows that resonant tunnelling trajectories are present owing to a certain type of residual stress---as discussed above---and can be counted using general arguments.

The RFOT theory predicts, then, that in a glass made by quenching near $T_\text{cr}$, a TLS should encompass 20 to 30 beads. For slower quenches, this figure should increase to reach about 150 to 300 beads near the glass transition on typical laboratory timescales of minutes to hours. In contrast, Mocanu et al.~\cite{MocanuNTLS} find that on the low-energy side, their bistable modes tend to involve at most a few particles. Only upon approaching energies relevant near the crossover, can configurations become quite collective, encompassing close to a hundred particles. In other words, both the energy trend and the absolute size for the extent of cooperativity seem to be in conflict between the two studies.

To sort this out, let us first get a quantitative sense for the bead size in the liquid mixture employed in Ref.~\cite{MocanuNTLS}. In the Mott-Gurney argument~\cite{MottGurneyLiquids, L_AP}, the entropy decrease that results from precluding particles to swap places is $\Delta S = k_B \ln[(z/N)^N] - k_B \ln(z^N/N!) = -N k_B$, or simply $-k_B$ per particle; $z$ is the one-particle partition function, which we deliberately fix so as to cleanly isolate the particle-exchange contribution to the entropy change from the other contributions. For a liquid mixture this entropy loss becomes, per particle, $\Delta S/N = k_B \{ \ln[(1/N)^N] - \ln(\prod_i 1/N_i!) \}/N = - k_B (1 - \sum_i x_i \ln x_i)$, where $x_i \equiv N_i/N$ is the mole fraction of species $i$ and one readily recognizes the expression for the mixing entropy $-\sum_i x_i \ln x_i$. For the (4 large):(1 small):(1 medium) particle mixture employed in Ref.~\cite{MocanuNTLS}, this mixing entropy is substantial, $\approx 0.87$. If one were to represent the mixture as a collection of chemically equivalent beads, a bead would then contain about $1/(1 + 0.87) < 1$ original particle. This can be contrasted with the bead count for actual substances. A bead usually contains at least a couple of atoms but can be as large as a naphthyl group~\cite{LW_soft}. 

This very large number of extra degrees of freedom---that would have to freeze out for a solid to form---suggests that vitrification in Lennard-Jones mixtures is accompanied by a variety of additional processes that are not manifested in actual liquids. A very recent, computation-intensive study on a binary mixture~\cite{PhysRevE.111.055415} suggests these special processes have to do with the formation of locally favoured structures and, concurrently, a depletion in the number of excitations in the liquid, owing to ensuing geometric frustration. With this in mind, one would be justified in inquiring whether those low-energy bistable modes in Ref.~\cite{MocanuNTLS}, which contained only several particles, have to do with competition among distinct locally favoured structures. If so, the modes would be peculiar to mixtures of the type studied in Ref.~\cite{MocanuNTLS}. 

\begin{figure}
\centering
\subfloat[Temperature dependence of the typical transition-state size
$N^\ddagger$ for the spatially-compact, $\alpha$-relaxations, for select values of the fragility index. The cooperativity size $N^* = 4 N^\ddagger$.]{%
\resizebox*{7cm}{!}  {\includegraphics{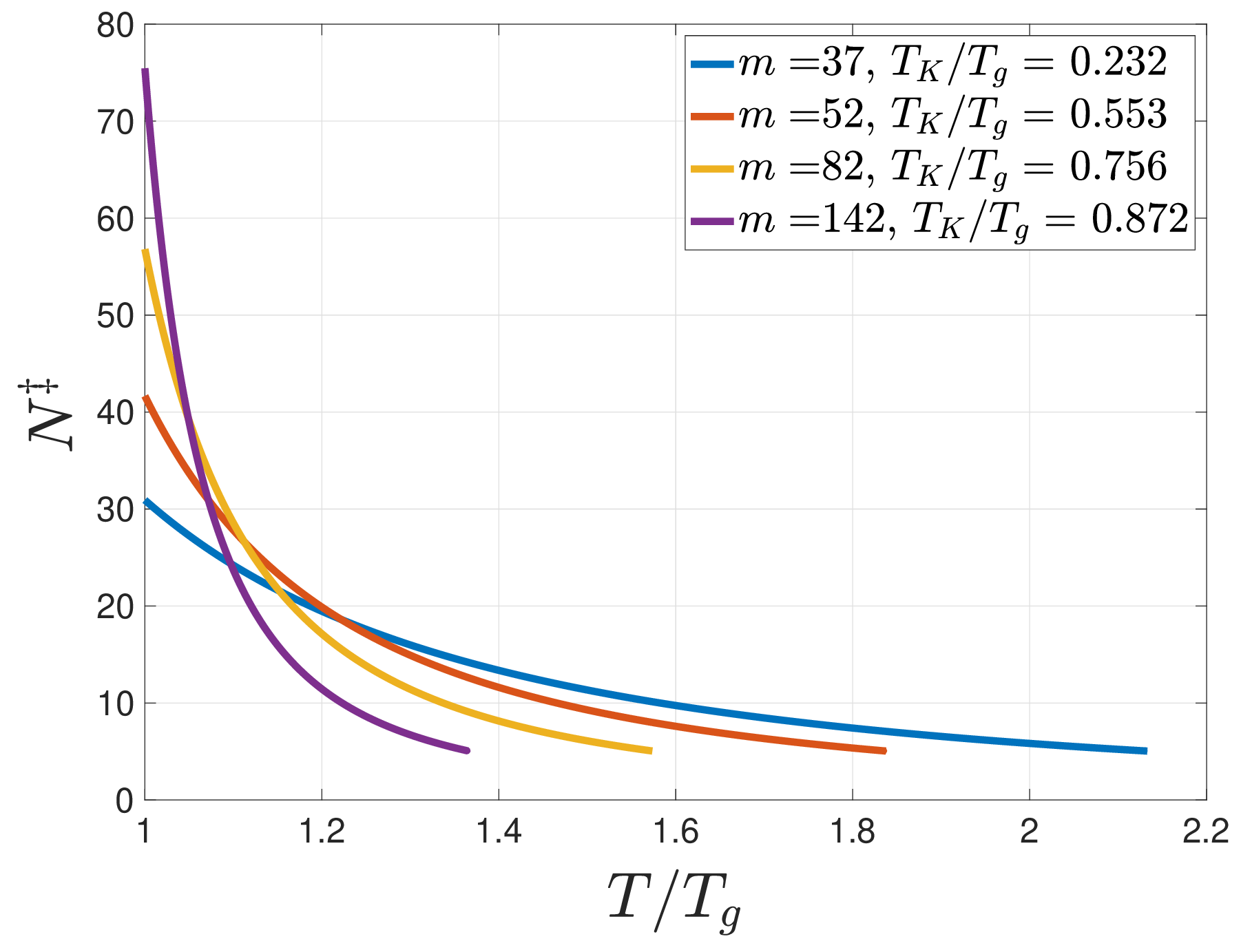}}}\hspace{5pt}
\subfloat[Temperature dependence of the typical string length $N_s$ for
the $\beta$-relaxations, for select values of the fragility index $m$.]{%
\resizebox*{7cm}{!}{\includegraphics{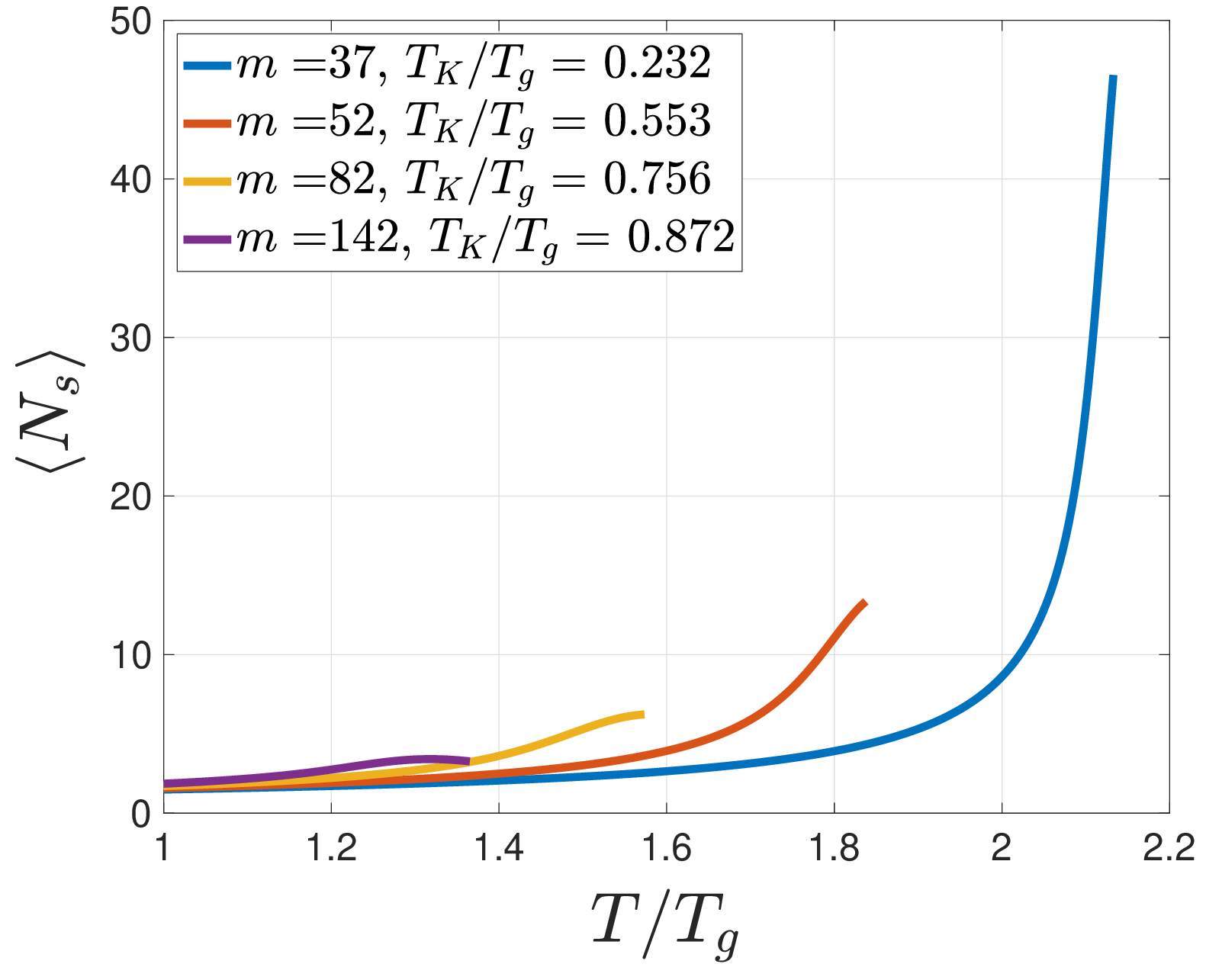}}}
\caption{Temperature dependence of the degree cooperativity for (a) $\alpha$-relaxations and (b) $\beta$-relaxation. Reprinted from Ref.~\cite{LWphoto} by permission.} \label{coopT}
\end{figure}

Another clue comes from the apparent increase in the cooperativity size with energy, in Ref.~\cite{MocanuNTLS}. The RFOT theory predicts~\cite{SSW, SWbeta, LWphoto}, that activated reconfigurations in glassy melts can be roughly classified into two distinct varieties. We have so far mentioned only the high-barrier variety, responsible for $\alpha$-relaxation. These reconfigurations are compact, their cooperativity increasing with cooling, as already mentioned. The other variety accounts for the $\beta$-relaxations. The corresponding relaxation processes are non-compact, string-like. For these, the cooperativity, in contrast, increases with temperature. We show both of these trends in Fig.~\ref{coopT}. It would be interesting to check whether those extended high-energy bistable modes reported in Ref.~\cite{MocanuNTLS} correspond to string-like modes.

Conversely, it appears that the preparation protocol used in Ref.~\cite{MocanuNTLS} would specifically prevent one from detecting bistable modes that could span the volume $\xi^3$. Indeed, the swap Monte Carlo is specifically designed to avoid transition states for {\em physically realizable ways} to perform $\alpha$-relaxation, see a discussion in Ref.~\cite{HLtime}. Consequently, the TLS search in Ref.~\cite{MocanuNTLS} is limited to structures---as obtained using swap Monte Carlo---each of which represents a free energy basin separated from other such basins by $\alpha$-relaxation barriers. This limitation seems to be consistent with a relatively narrow distribution of the energies of the detected inherent structures. Indeed, one expects two standard deviations (per particle, and setting $k_B =1$): $2 \delta E/N \approx 2 T \sqrt{\Delta c_p/N}$, where $\Delta c_p$ is now per particle, not bead. For the parameters in Ref.~\cite{MocanuNTLS}  this yields $2 \delta E/N \approx 0.028 \cdot \sqrt{\Delta c_p}$, i.e., likely greater than $0.028$, in view of the earlier discussion of the bead count and the substance being fragile. The energy range for inherent structures reported in Ref.~\cite{MocanuNTLS} is, however, substantially less: $0.0076$.

\subsection{Matured, not aged?}

Like the two systems considered above, the ancient amber from Ref.~\cite{PhysRevLett.112.165901} {\em also} represents a solid that is stabilized to a low enthalpy state unavailable to conventional quenching protocols. The degree of stabilization appears to be somewhat less than that of the most stable deposited films but is still substantial: The melting temperature of the matured amber is about 20K above the melting temperature of the rejuvenated sample. Yet in contrast with those two systems, the stabilization in amber took place under conditions that were at no point equilibrium or ever resulted in an equilibrium configuration. Instead, a resin that had already vitrified subsequently underwent chemical processes of polymerization and cross-linking. To simplify reasoning, we will assume that the ambient conditions under which these chemical processes took place were steady.

\begin{figure}
\centering
\subfloat[Sketch of the temperature dependences of the number density $\rho$ and configurational entropy $s_c$. The blue lines correspond to the equilibrium equation of state, while the magenta lines depict what happens if the liquid vitrifies. The cyan arrows show what happens (locally) during initial ageing events. Thereby the (local) fictive temperature relaxes to some value $T_\text{rlx}$, such that $T_\text{amb} < T_\text{rlx} < T_g$, where $T_\text{amb}$ is the ambient temperature. $T_\text{rlx} $ approaches $T_\text{amb}$ for shallow quenches~\cite{L_AP}.]{%
\resizebox*{6cm}{!}  {\includegraphics{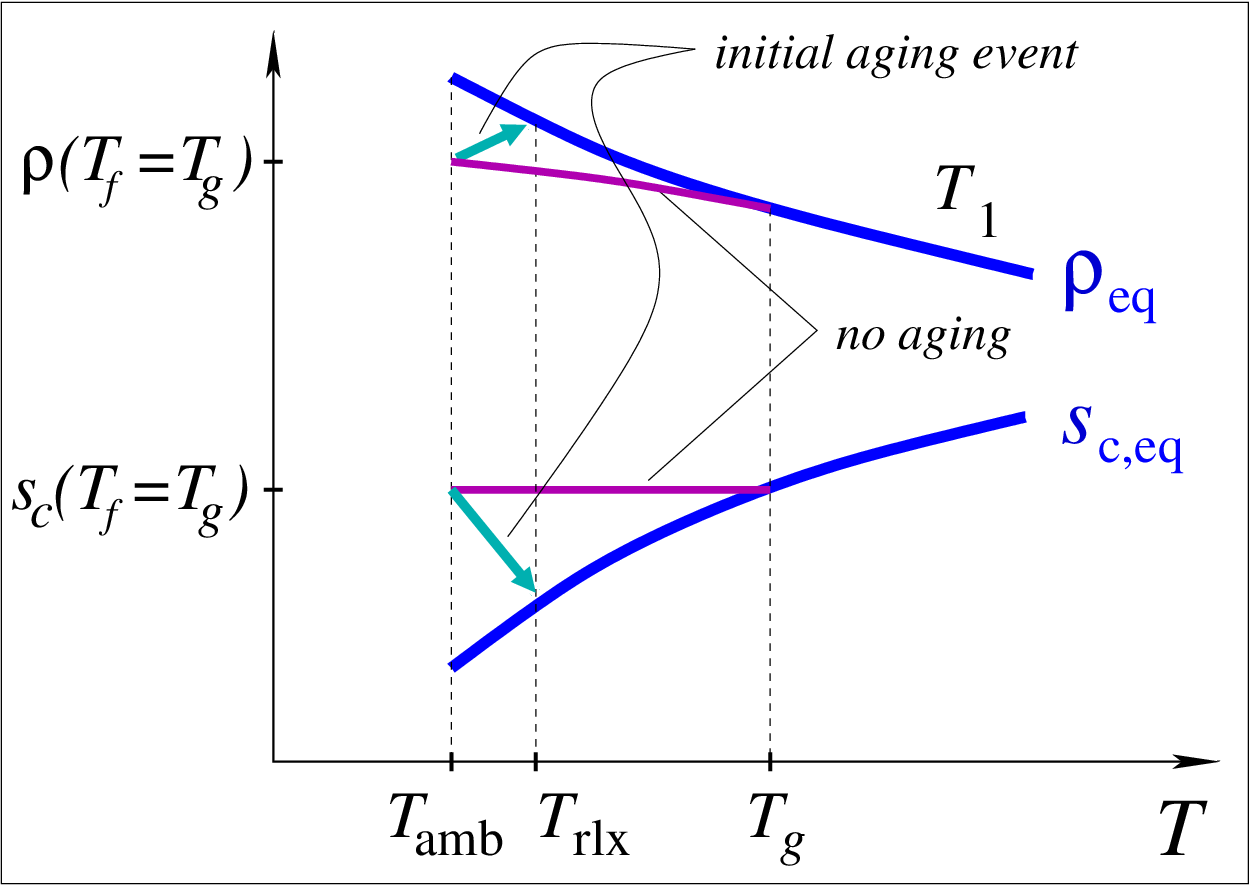}}}\hspace{5pt}
\subfloat[The off-equilibrium diagram of a liquid in the $(\rho, p^{-1})$ plane, where $p$ is the pressure. Subscript ``K'' signifies the putative Kauzmann state. Reprinted from ~\cite{LWjamming} under a Creative Commons CC BY 4.0 International License.]{%
\resizebox*{6cm}{!}{\includegraphics{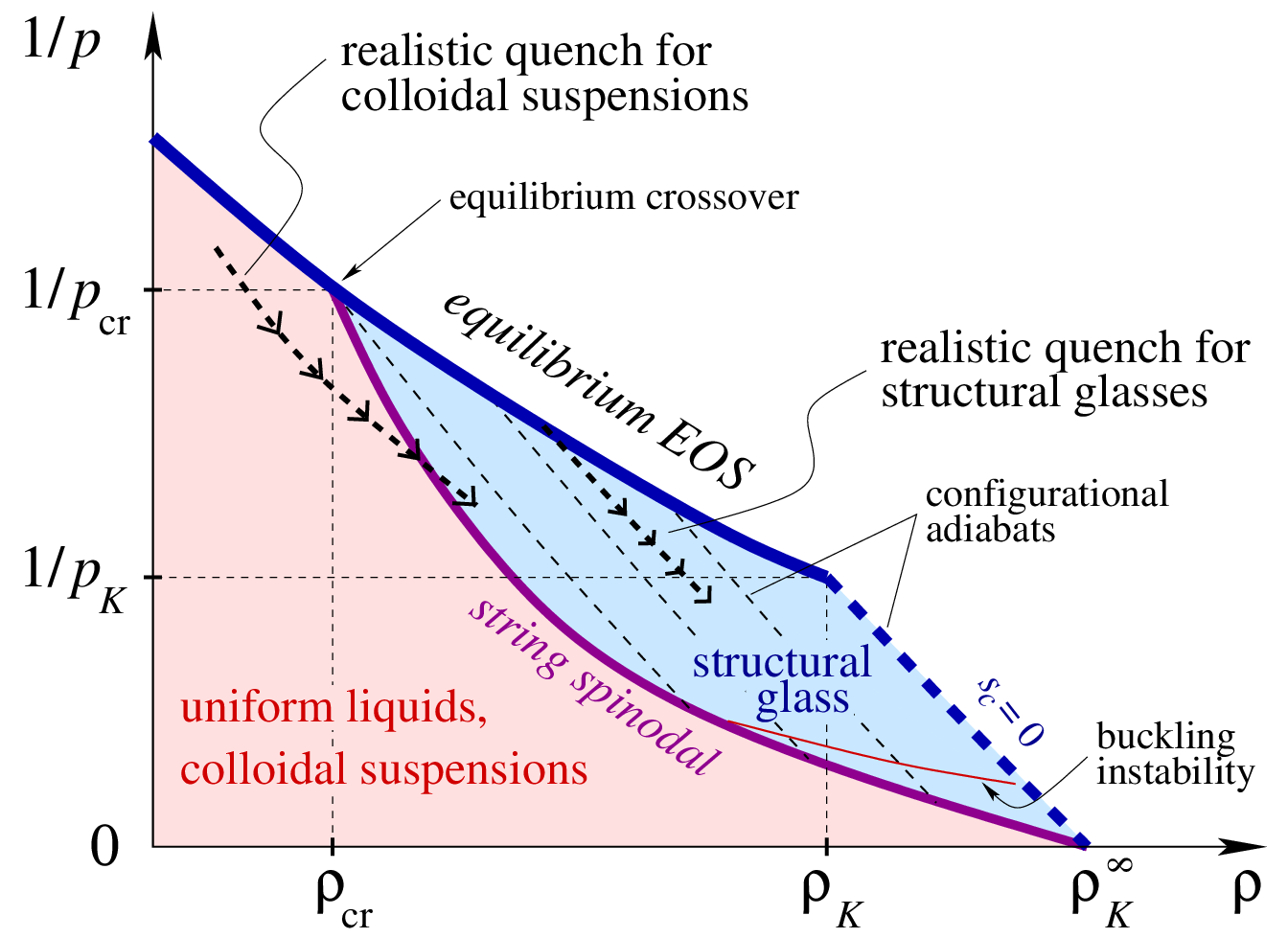}}}
\caption{Diagrams juxtaposing equilibrium and off-equilibrium behaviours of a liquid.} \label{jamming}
\end{figure}

We first note that the pristine, matured amber is about 2\% denser than the same material after annealing~\cite{PhysRevLett.112.165901}. The contraction could have resulted from a combination of two distinct processes. One is a contraction caused by the polymerization-induced increase in bonding. The contraction is largely uniform, even though some non-affine motions are likely to occur alongside, too; these would be reversible. The other process, in contrast, has to do with irreversible, strongly anharmonic structural reconfigurations we associate with ageing~\cite{LW_aging}. Ageing takes place because the structure of a glass is representative not of the ambient conditions, but of the conditions under which vitrification took place. One often uses the approximate concepts of ``fictive temperature'' and ``fictive pressure'' to specify those conditions. Whenever the fictive values of temperature and pressure differ from their ambient values, there is a driving force toward ageing. These notions are illustrated in Fig.~\ref{jamming}(a), where we graphically illustrate that the configurational entropy and the density each exhibit a slope discontinuity at the glass transition, caused by the freezing-out of the configurational subset of the motions. For any glass formed above the ambient temperature, ageing will locally drive the glass toward lower values of the configurational entropy because the configurational entropy is a monotonically increasing function of temperature.

It stands to reason that amber would age. The melting temperature of amber ranges from $415$~K for rejuvenated samples to $440$~K for pristine matured samples~\cite{PhysRevLett.112.165901}. A {\em non-polymeric} glass would exhibit early aging events on time scales $\sim 10^6 \ldots 10^7$~hrs, near $300$~K. Indeed, the $\alpha$-relaxation barrier is about $37 k_B T_g$~\cite{XW, L_AP}, and $\exp[37 \cdot (415/300-1)] \sim 10^6$, $\exp[37 \cdot (440/300-1)] \sim 10^7$. In contrast, the age of the amber is $10^8$ years, i.e., five or six orders of magnitude longer. P\'erez-Casta\~neda et al.~\cite{PhysRevLett.112.165901} do view the increase in stability of the matured amber as a result of ageing. In turn, this can be interpreted as a decrease in the fictive temperature $T_f$ of the sample. Using extrapolations of enthalpy, the inferred decrease in $T_f$ is $32$~K~\cite{PhysRevLett.112.165901}. This would, however, conflict with the RFOT theory, in view of the quadratic scaling $1/N^* \propto s_c^2$, c.f. Eq.~(\ref{FNsc}). Amber is considered a fragile substance~\cite{doi:10.1021/acs.jpclett.9b00003}. Lowering of the vitrification temperature of a fragile substance by nearly 8\% should correspond to a noticeable decrease in the configurational entropy, 20\%-to-30\%. The resulting depletion of the TLSs would be comparable to 50\%. But amber barely shows any depletion. 

In attempting to resolve this discrepancy, one may ask, for the sake of argument, what would have happened if no ageing occurred whatsoever. This question is reasonable in view of old experiments of Kovacs and collaborators~\cite{Kovacs1977}, who have directly shown that glassy polymers exhibit memory effects. Hereby, one follows the time evolution of the volume of a polymer quenched from a relatively high temperature $T_1$ to a relatively low temperature $T_3 < T_1$ and then warmed up to an intermediate temperature $T_2$ ($T_1 > T_2 > T_3$). This time evolution turns out to be non-monotonic so as to partially ``retrace'' the trajectory for the direct $T_1 \rightarrow T_2$ process. We suppose, then, that the polymerization caused by maturation introduces an additional constraint to activated transitions in the glass. This constraint would effectively add to the mismatch penalty $\Gamma(N)$ for {\em new} reconfigurations, since polymerization occurred after vitrification, when the structure was already set. The increase in bonding would lead to uniform compression, as already mentioned. 

Effects of compression on a vitrified sample were considered by Lubchenko and Wolynes in Ref.~\cite{LWjamming}. These authors put forth an extended phase diagram for liquid-to-solid transitions that covers a variety of preparation protocols for making off-equilibrium solids, ranging from jammed particulate assemblies to vitreous solids held together by molecular forces. The diagram is shown in Fig.~\ref{jamming}(b). (In the diagram the glass transition is also induced by compression, not cooling, but this is immaterial.) Of interest here is the protocol labelled ``realistic quench for structural glasses,'' which shows what happens when a bulk-quenched glass is subjected to hydrostatic pressure, at fixed temperature. The sample follows one of the configurational adiabats until it begins to irreversibly relax toward structures that are denser and, at the same time, correspond to lower values of the configurational entropy. The barrier for such reconfigurations is, however, very tall unless the external pressure is very high---of order gigapascals---that would typically cause electronic transitions, in the first place. Given that the bulk modulus of the amber samples is about $7$~GPa~\cite{PhysRevLett.112.165901}, the excess built-in pressure needed to cause a 2\% contraction is about $140$ MPa or so, which appears to be below what could cause substantial ageing on typical laboratory time scales, according to Ref.~\cite{LWjamming}. 

The simplest realization of the RFOT theory~\cite{XW} predicts that the laboratory glass transition on the minute-to-hour scale corresponds, nearly universally to $s_c \simeq 0.8~k_B$ per bead; improved estimates yield system-dependent corrections~\cite{RWLbarrier, LRactivated}. This implies that the glass transition temperature and pressure, respectively, are tied through the relation  $s_c(p_g, T_g) = \text{const}$. A quick look at Fig.~\ref{jamming}(a) will readily convince one that the slope $dT_g/dp_g$ of the ``phase boundary" should be positive. Consequently, if the configurational entropy of the matured, compressed amber did not change much from the moment of vitrification, the (now-pressurized) amber should melt at a temperature higher than would a sample that has been rejuvenated and subsequently re-vitrified. To quantify this, we turn to Roland et al.~\cite{Roland_2005}, who report slopes $dT_g/dp_g$ in the range 110-380 K/GPa for a variety of substances, mostly organic and/or polymeric. For the amount of built-in pressure in the matured amber samples from Ref.~\cite{PhysRevLett.112.165901}, $140$~MPa, this would imply an increase in the melting temperature ranging from 15 to 50K. The calorimetry data in Fig.~2 of Ref.~\cite{PhysRevLett.112.165901} fall within this range and, thus, would be consistent with little, if any, decrease in the configurational entropy. Consequently, in view of Eqs.~(\ref{nTLS})-(\ref{FNsc}), the TLS density in amber would not change much during maturation, thus resolving the contradiction.

\section{Summary and outlook}

\label{summary}

It is truly remarkable that relics of motions responsible for melting processes should survive at cryogenic temperatures. Indeed, except when protected by symmetry, low-energy degrees of freedom tend to eventually develop a gap---owing to interactions---and freeze out at sufficiently low temperatures: Electrons in metals form Cooper pairs and then Bose-condense; protons in ice eventually settle into a lattice despite the vast number of available configurations~\cite{stillinger2015energy}; low-energy tunnelling motions in quantum solids are expected to develop a superfluid component~\cite{RevModPhys.84.759}. Yet the spectrum of the two-level systems extends to the lowest energies accessed so far, as if they were truly massless excitations like the phonons. Lubchenko and Wolynes~\cite{LW} have argued that this abundance of low-energy excitations stems from the huge density of states with continuously distributed energies available to an equilibrium glassy melt. The number of these states scales exponentially with the system size, $e^{N s_c/k_B}$ ($s_c \sim 1 k_B$), as would the number of states for the phonons, $\sim e^{N s_\text{vib}/k_B}$. Although the vibrational entropy $s_\text{vib}$ is greater than the configurational entropy $s_c$, both entropies are of the order $10^0 k_B$, near melting. The two degrees of freedom become largely decoupled once the liquid melt is cooled below the crossover from collisional to activated transport, centred at $T_\text{cr}$, but the two never truly divorce from each other~\cite{FreemanAnderson}: The Boson peak---which is the parent excitation for the TLS~\cite{LW_BP, LW_RMP}---prevents the phonons from becoming quasi-particles down to THz frequencies, while at frequencies below, a phonon will travel for a hundred wavelengths or so before it is resonantly absorbed by a two-level system. (At this point, the caloric content of the glass is dominated by the TLSs.) To put this in perspective, a phonon in a periodic crystal would have no trouble traversing the {\em whole sample} at these temperatures.

The exponentially large number of metastable structural configurations sampled by a glassy melt can be thought of as a library of states~\cite{LW_aging}. Once the melt freezes, the contents of the library become settled---aside from some ageing---thus documenting the preparation protocol; the two-level systems can be poetically thought of as the library's catalogue, but one that has been relegated to a frigid basement. 

Several recent studies---of actual systems and computational models---confirm that the two-level systems are, indeed, a sensitive indicator of the preparation history of the sample, even though the preparation itself was at temperatures that are two orders of magnitude higher than the energies where the TLSs begin to dominate the thermal properties of the sample. In all cases, the samples were substantially stabilized in terms of enthalpy, but the presence of stabilization alone is insufficient to predict whether structural changes took place that could affect the density of states of the TLSs.  

The structure in ultrastable films made by vapor deposition appears to lack long range order yet it exhibits a great deal of {\em local} ordering. It seems likely that the molecules in ultrastable films and quenched melts, respectively, pack sufficiently differently as to require nucleation to effect transitions between the two structures, similarly to distinct phases separated by a discontinuous transition. The more efficient packing in the film, we argued, underlies a substantial decrease in the residual stress and configurational-entropic content of the substance. This could account for the apparent decrease in the density of states of the TLSs in very stable films. 

If local ordering were to play a significant role in this effect, one might be justified in deliberately tuning the degree and type of ordering through molecular engineering. For instance, there is a class of glass-formers that exhibit a smectic type of local ordering~\cite{doi:10.1021/acs.jpcb.5c03603}. It would be interesting to measure low-$T$ properties of such smectic glasses. It appears that in order for ultrastable films to be thermodynamically competitive with quenched melts, their vibrational entropy at temperatures near melting must be greater. This should be testable using a combination of spectroscopy and calorimetry. On the theory front, the state-counting argument implemented through the library construction~\cite{LW_aging, LW} can be presumably generalized to include local ordering and/or anisotropic beads. Oddly enough, Prince Rupert's name is associated with a relevant problem in geometry~\cite{steininger2025convexpolyhedronrupertsproperty}. 

Unlike the ultrastable films, the enthalpically-stable glassy mixtures generated computationally by Mocanu et al.~\cite{MocanuNTLS} may well reside on the equilibrium equation of state (EOS) line. Candidate TLS modes in the form of bistable energy profiles are readily identifiable in these samples. The success rate of finding such profiles decreases with stabilization, consistent with RFOT-based predictions on the relation between the density of structural states and the TLS number~\cite{LW}. However the degree of the cooperativity of the candidate modes and its trend with energy are in conflict with the RFOT-based predictions. This seeming conflict may be resolved by noting that the search algorithms in Ref.~\cite{MocanuNTLS}  are limited to motions that specifically avoid transition-state configurations for $\alpha$-relaxations. In contradistinction, Lubchenko and Wolynes~\cite{LW, LW_RMP} argued that the TLSs correspond to resonant modes that contribute to the low-barrier flank of $\alpha$-relaxations. Conversely, the cooperativity trends exhibited by the bistable modes in study~\cite{MocanuNTLS} seem to match the characteristics of non-compact, stringy excitations that have been argued to contribute to the $\beta$-relaxations in the melt~\cite{SWbeta, LWphoto}. This notion can be tested by examining the shapes of the cooperative reconfigurations found in Ref.~\cite{MocanuNTLS}. In addition, it would be instructive to check whether some of the less collective modes found in the latter study correspond to transitions between locally favoured structures that are peculiar to model LJ mixtures. 

While ageing leads to enthalpic stability, the converse is not necessarily true. Amber samples matured over geological times appear to be an example of such a situation. Inspired by old experiments of Kovacs and collaborators on memory effects in polymers~\cite{Kovacs1977}, we have considered a possibility that the chemical processes of polymerization and cross-linking amounted to shortening and stabilizing the bonds---through an added, uniform compression of the sample---but have not significantly modified the structure. Moreover, the added linkage due to the strengthened bonds helped preserve the structure. Thus the density of states that was frozen-in at the glass transition in the fresh resin remained frozen-in as the sample was being compressed by those chemical forces. This notion is certainly speculative but it is, nonetheless, consistent with known relationships between the temperature and pressure of the glass transition, and could potentially explain why matured ambers do not show an appreciable depletion of the two-level systems. 

The temperature variations in Kovacs' experiments are too small to have much effect on the TLS density of states, but perhaps a similar experiment performed using {\em pressure} quenches would be more informative. For instance, Yue et al.~\cite{10.1063/1.2719194} performed quenches (of non-polymeric substances) at pressures as high as 500 MPa, which is substantially higher than our estimate for the effective increase in pressure caused by maturation. In the latter experiment, glasses formed by pressurizing followed by cooling, exhibited a substantial enthalpy overshoot during melting. Thereby, the height of the DSC peak increased with the amount of pressure used to prepare the sample, while the melting temperature itself changed little. 

In conclusion, since the amount of the two level systems is directly tied to the configurational-entropy content of the solid, there are two ways to manipulate their amount. One way is to effect a transition to a solid with a distinctly different spectrum of aperiodic configurations. The other possibility is to vary the quench rate of the glassy melt. We expect that the density of the TLSs will increase by an order of magnitude or so for hyper-quenched glasses that fell out of equilibrium near the crossover temperature, relative to glasses made using leisurely cooling rates. Since the rate of thermal quenching is quite limited by the thermal conductivity of the sample, quenching via compression may be the more promising way to go. In contrast with simulations, which strive to reach the ``glass ceiling'' of long relaxation times~\cite{doi:10.1073/pnas.1706860114}, experimental quenches face the opposite challenge of reaching the ``glass floor'' of fast quenches. By seeing directly through this glass floor, even if darkly~\cite{YuLeggett, LSbinoculars}, we could perhaps observe the two-level systems in greater numbers.  

\section*{Funding}
I gratefully acknowledge the support by a grant from the Texas Center for Superconductivity at the University of Houston, the NSF Grant CHE-1956389, and the Welch Foundation Grant E-1765.

\bibliographystyle{tfq}
\bibliography{lowT}

\end{document}